\documentclass[adp,a4paper,
]{w-art}
\usepackage{times,cite,w-thm,amsfonts}
\theoremstyle{plain}
\newcommand{\Tr}{{\rm Tr}}

\theoremstyle{definition}

\usepackage[]{graphicx,pstricks,pst-node}
\usepackage{amsmath,amssymb}
\numberwithin{equation}{section}

\newcommand{\hh}{\hat{H}}
\newcommand{\ha}{\hat{a}}
\newcommand{\hb}{\hat{b}}
\newcommand{\hc}{\hat{c}}
\newcommand{\hs}{\hat{s}}

\newcommand{\rmi}{{\rm i}}

\begin{document}
\DOIsuffix{theDOIsuffix}
%%
%% issueinfo for the header line
\Volume{16}
\Month{01}
\Year{2007}
%%
%%    First and last pagenumber of the article. If the option
%%    'autolastpage' is set (default) the second argument may be left empty.
\pagespan{1}{}
%%
%%    Dates will be filled in by the publisher. The 'reviseddate' and
%%    'dateposted' (Published online) entry may be left empty.
\Receiveddate{XXXX}
\Reviseddate{XXXX}
\Accepteddate{XXXX}
\Dateposted{XXXX}
\keywords{Quantum time evolution, non-Hermitian quantum mechanics, 
dephasing, relaxation}

%% \pretitle{Editor's Choice}

%% We have a short and a long form for the title. The short form
%% (optional argument) goes into the running head.

\title[Particle injection into a chain]%
% {Time evolution 
% for particle injection into a chain}
{Particle injection into a chain: decoherence versus relaxation for Hermitian and non-Hermitian dynamics}

\author[F.\ Gebhard]{F.\ Gebhard\inst{1,}%
\footnote{Corresponding author\quad  
E-mail:~\textsf{\scriptsize florian.gebhard@physik.uni-marburg.de}, Phone: 
+49\,6421\,282\,1318, Fax: +49\,6421\,282\,4511}}
%%
%%    Information for the second author
\author[K.\ zu M\"unster]{K.\ zu M\"unster\inst{1}}
\author[J.\ Ren]{J.\ Ren\inst{2}}
\author[N.\ Sedlmayr]{N.\ Sedlmayr\inst{2}}
\author[J.\ Sirker]{J.\ Sirker\inst{2}}%
\author[B.\ Ziebarth]{B.\ Ziebarth\inst{1}}
\address[\inst{1}]{Dept.\ of Physics, 
Philipps-Universit\"at Marburg,
D-35032 Marburg, Germany }
\address[\inst{2}]{Dept.\ of Physics and Research Center OPTIMAS, 
TU Kaiserslautern, D-67663 Kaiserslautern, Germany}

\begin{abstract}
  We investigate a model system for the injection of fermionic
  particles from filled source sites into an empty chain. We study the
  ensuing dynamics for Hermitian as well as for non-Hermitian time
  evolution where the particles cannot return to the bath sites
  (quantum ratchet).  A non-homogeneous hybridization between bath and
  chain sites permits transient currents in the chain.
  Non-interacting particles show decoherence in the thermodynamic
  limit: the average particle number and the average current density
  in the chain become stationary for long times, whereas the
  single-particle density matrix displays large fluctuations around
  its mean value. Using the numerical time-dependent density-matrix
  renormalization group ($t$-DMRG) method we demonstrate, on the other
  hand, that sizable density-density interactions between the
  particles introduce relaxation which is by orders of magnitudes
  faster than the decoherence processes.
\end{abstract}
\maketitle  

\section{Introduction}

Newton's equations describe the deterministic time evolution of a
given initial state of classical particles. In principle, the particle
velocities and positions are known for all times.  In practice, in
isolated systems and for long times, macroscopic observables which
involve the averages over many particles can be calculated equally
accurately starting from thermodynamic considerations.  The
`thermalization' of macroscopic systems is one of the building
principles of statistical mechanics. According to statistical
mechanics, expectation values for macroscopic observables should, for
long times, become independent of time and, moreover, become equal to
their statistical average; for a recent example of the relaxation of
classical particles in a one-dimensional box, see
Ref.~\cite{MuensterGebhard}. The `thermalization principle' equally
applies to quantum mechanics where the Schr\"odinger equation
describes the deterministic evolution of a given initial state in
time. Recently, the question under which conditions thermalization is
possible in closed quantum systems has also been studied, both
experimentally and theoretically
\cite{KinoshitaWenger,HofferberthLesanovsky,RigolDunjko}. According to
the eigenstate thermalization hypothesis, each generic state of a
closed quantum system already contains a thermal state which is
revealed during unitary time evolution due to dephasing and relaxation
processes. However, proofs for this hypothesis have so far only been
obtained in very specific settings \cite{Deutsch,Srednicki}; its
applicability for generic quantum systems remains a matter of current
research.

In this work, we are interested in the quantum-mechanical injection of
particles into a chain. For this toy model we can study the approach
to a steady state analytically and numerically. The time evolution of
prepared initial states in one-dimensional quantum systems such as our
toy model can be studied experimentally using cold atomic gases on
optical lattices confined to one dimensional potentials
\cite{KinoshitaWenger,HofferberthLesanovsky,ott,strohmaier}. These
systems are to a very high degree isolated from the environment and
are therefore ideally suited to study non-equilibrium dynamics in
closed quantum systems. Besides the Hermitian time evolution we also
address the non-Hermitian case for our toy model where the chain
particles cannot return to the bath sites (a `quantum
ratchet'~\cite{Haenggi,adpHaenggi}). In general, effective
non-Hermitian Hamiltonians are naturally obtained if one simulates a
dissipative environment for a quantum model in the form of a Lindblad
equation \cite{CarloBenenti}. An effective non-Hermitian Hamiltonian
has also been used to study the depinning of flux lines in
superconductors \cite{HatanoNelson}. Interesting from a fundamental
point of view are, in particular, so-called pseudo-Hermitian
Hamiltonians---including Hamiltonians having
$\mathcal{PT}$-symmetry---which have real spectra and therefore offer
a natural framework to extend standard quantum mechanics
\cite{Mostafazadeh,BenderBoettcher,Bender_review}.

In Sect.~\ref{sec:nonhermitian} we recapitulate the Schr\"odinger
equation for time evolution in quantum mechanics. We treat both
cases of Hermitian and non-Hermitian Hamilton operators.  In
Sect.~\ref{sec:particlemodel} we introduce our model system for
spinless fermions. We consider a chain of sites which is initially
empty and is coupled to initially filled bath sites. At finite times,
the hybridization introduces particles in the chain where they can
move between neighboring sites. The hybridizations can be homogenous
or inhomogeneous; in the latter case they induce a transient current
in the chain.

In Sect.~\ref{sec:noninterating} we discuss the particle density, the
single-particle density matrix, and the current density for
non-interacting fermions. In the thermodynamic limit, the particle and
current densities become constant for large times because the
contributions of individual modes interfere destructively
(`decoherence').  The oscillations in the single-particle density
matrix on the other hand reveal the absence of relaxation for
non-interacting particles. Furthermore, we investigate the influence
of the ratchet condition on the chain dynamics.  In
Sect.~\ref{sec:interacting} we include a density-density interaction
between the particles.  We use the time-dependent density-matrix
renormalization group ($t$-DMRG) method to evolve the
quantum-mechanical states in time.  The interaction between the
particles induces a fast relaxation to stationary distributions for
the particle density, single-particle density matrix, and current.
For our model system, with interactions of the order of the bandwidth,
we find that the relaxation time scale is two orders of magnitude
shorter than the time scale for dephasing.

A summary and conclusions, Sect.~\ref{sec:penultimate} 
% and Sect.~\ref{sec:final}
, close our presentation.  Calculations for the
asymmetric chain filling in the ratchet case are deferred to
appendix~\ref{app:D}.

\section{Hermitian and non-Hermitian time evolution}
\label{sec:nonhermitian}

We begin our discussion with the Schr\"odinger equation
which defines the time evolution for Hermitian and non-Hermitian 
Hamilton operators. As in standard quantum mechanics,
expectation values are to be taken using the 
time-dependent wave function. The von-Neumann equation of motion
of the corresponding statistical operator can be cast into the
general form of a Kossakowski--Lindblad equation. 

\subsection{Schr\"odinger equation}

We consider a finite-dimensional Hilbert space $\mathbb{V}$.
Its elements are the states $|\psi\rangle$.
We use the
standard scalar product $\langle \phi|\psi\rangle\in \mathbb{C}$
where $\langle \phi |$ is the shorthand notation for the linear functionals
$\langle \phi| \cdot \rangle$ which form the dual space $\mathbb{V}^*$
of our Hilbert space.
For an operator $\hat{O}$ which acts in $\mathbb{V}$,
its adjoint $\hat{O}^{\dagger}$ is defined by
$\langle \phi | \hat{O}^{\dagger}| \psi \rangle =
\langle \psi | \hat{O} |\phi\rangle^*$
for all $\langle \phi | \in \mathbb{V}^*$ and $|\psi\rangle\in \mathbb{V}$
where $z^*$ is the complex conjugate of $z\in \mathbb{C}$.

For time $t\geq 0$, the time evolution 
of a state $|\psi\rangle$ and of its adjoint state $\langle \psi|$ 
is given by
\begin{equation}
|\psi(t)\rangle= \exp\left(-\rmi\hat{H} t\right) |\psi\rangle 
\quad , \quad 
\langle \psi(t)| = \langle \psi| 
\exp\left(\rmi\hat{H}^{\dagger} t\right) 
\; ,
\label{eq:schroedinger}
\end{equation}
where $\hat{H}$ is a time-independent Hamilton operator and
$\hat{H}^{\dagger}$ is its adjoint.

In the following we shall assume that the spectrum of $\hat{H}$ is real,
i.e., the eigenstates of $\hat{H}$ obey
\begin{equation}
\hat{H} |E_n\rangle = E_n |E_n\rangle \quad (n=1,2,\ldots)
\end{equation}
with $E_n\in \mathbb{R}$. As shown in Ref.~\cite{Mostafazadeh} this
implies that $\hat{H}$ is pseudo-Hermitian and $\hat{H}^{\dagger}$ has
the same spectrum,
\begin{equation}
\hat{H}^{\dagger} |\overline{E_n}\rangle = E_n |\overline{E_n}\rangle 
\quad (n=1,2,\ldots) \; .
\label{eq:eigenstatesofHdagger}
\end{equation}
Eq.~(\ref{eq:eigenstatesofHdagger}) implies that
there is a left eigenvector $\langle \overline{E_n} |$ of $\hat{H}$ 
for every right eigenvector $|E_n\rangle$ 
of $\hat{H}$ with the same eigenvalue $E_n$.

Note that $\langle \overline{E_n} |= \left(|E_n\rangle\right)^{\dagger}$
only if $\hat{H}$ is Hermitian, $\hat{H}^{\dagger}=\hat{H}$.
The eigenstates of $\hat{H}$ and $\hat{H}^{\dagger}$ for different energies
are orthogonal 
$\langle \overline{E_n} |E_m\rangle=0$ if $E_n\neq E_m$.
Using proper linear combinations of eigenstates for degenerate
eigenvalues we can write for the Hamiltonian, its adjoint, 
and the unit operator
\begin{equation}
\hat{H}= \sum_n E_n |E_n\rangle \langle \overline{E_n} |
\quad , \quad
\hat{H}^{\dagger}= \sum_n E_n |\overline{E_n}\rangle \langle E_n | 
\quad , \quad
\hat{1}= \sum_n |\overline{E_n}\rangle \langle E_n | 
= \sum_n |E_n\rangle \langle \overline{E_n} |=\hat{1}^{\dagger}\; ,
\end{equation}
where $\langle \overline{E}_n  |E_n\rangle=1$ thus forming a
biorthonormal set of eigenvectors. The Hamiltonian and its adjoint obey
the (pseudo-)Hermiticity condition~\cite{Mostafazadeh}
\begin{equation}
\hat{\eta} \hat{H}= \hat{H}^{\dagger} \hat{\eta}
\quad , \quad
\hat{\eta}= \sum_n |\overline{E_n}\rangle \langle \overline{E_n} |
=\hat{\eta}^{\dagger}  
\quad , \quad
\hat{\eta}^{-1}=\sum_n |E_n\rangle \langle E_n| \; .
\end{equation}
Note that all definitions reduce to the standard expressions
in the case of Hermitian time evolution with $\hat{H}=\hat{H}^{\dagger}$
and $\hat{\eta}=\hat{1}$.

\subsection{Measurements}

In standard quantum mechanics, the expectation value for an operator $\hat{O}$
is given by
\begin{equation}
\langle \hat{O} \rangle (t) =
\frac{\langle \psi(t)| \hat{O} | \psi(t)\rangle}%
{\langle \psi(t)| \psi(t)\rangle} \; .
\end{equation}
For pseudo-Hermitian time evolution, this definition remains unchanged.
It is important to note that the time evolution
described by~(\ref{eq:schroedinger}) is not unitary 
so that the norm $\langle \psi (t) | \psi(t)\rangle$ is not conserved
and must be considered separately.

As a consequence, the equation of motion for the statistical operator
changes its form. The statistical operator 
for the pure state $|\psi(t)\rangle$ is defined by
\begin{equation}
\hat{\rho}(t) = 
\frac{|\psi(t)\rangle\langle \psi(t)|}{\langle \psi (t)|\psi(t)\rangle} \;.
\end{equation}
With the help of the statistical operator, 
expectation values can be expressed in the form
\begin{equation}
\langle \hat{O}\rangle(t) = \Tr \bigl(\hat{\rho}(t)\hat{O}\bigr)\; .
\end{equation}
where $\Tr$ denotes the trace which implies the sum over 
the expectation values of an orthonormal basis set.
The statistical operator obeys $\hat{\rho}^{\dagger}=\hat{\rho}$,
$\hat{\rho}^2=\hat{\rho}$ and $\Tr\hat{\rho}=1$. 
Its time dependence is given by the equation of motion
(generalized von Neumann equation)
\begin{equation}
\rmi\frac{{\rm d}}{{\rm d} t} \hat{\rho}(t) =
\hat{H}\hat{\rho}(t)-\hat{\rho}(t)\hat{H}^{\dagger}
+\hat{\rho}(t)\langle \hat{H}^{\dagger}-\hat{H}\rangle(t) \; .
\end{equation}
When we split the Hamiltonian into a Hermitian part 
$\hat{H}_0=\hat{H}_0^{\dagger}$ and a non-Hermitian perturbation 
$\hat{V}\neq \hat{V}^{\dagger}$ we can equivalently write
\begin{equation}
\frac{{\rm d}}{{\rm d} t} \hat{\rho}(t) =-\rmi
\left[ \hat{H}_0,\hat{\rho}(t)\right]_{-}
-\rmi\left( \bigl(\hat{V}-\langle\hat{V}\rangle(t)\bigr) \hat{\rho}(t)
-\hat{\rho}(t)\bigl(\hat{V}^{\dagger}-\langle\hat{V}^{\dagger}\rangle(t)\bigr)
\right)\; .
\end{equation}
Apparently, the generalized von-Neumann equation takes the form of
the Kossakowski--Lindblad equation~\cite{Kossakowski,Lindblad}.

\section{Particle injection into a chain}
\label{sec:particlemodel}

As an example we study the injection of fermionic particles from
filled source sites into an empty chain. We start with a translationally
invariant coupling of the source and bath sites. Next, we consider an
inhomogeneous coupling which permits us to generate a transient
current density in the chain. For the case of non-interacting
fermions, analytical expressions can be worked out for the particle
density, the single-particle density matrix, and the current density.
For interacting fermions, we evolve the states in time with the
density-matrix renormalization group ($t$-DMRG) method.

\subsection{Hamilton operator}

We consider a finite chain with $L$~sites on which spinless fermions
can move between neighboring sites,
\begin{equation}
\hat{T}=-J \sum_{l=1}^{L-1} 
\left(
\hat{c}_{l+1}^{\dagger}\hat{c}_{l}+\hat{c}_{l}^{\dagger}\hat{c}_{l+1}
\right) \; ,
\end{equation}
where $\hat{c}_{l}^{\dagger}$ ($\hat{c}_{l}$) creates (annihilates) a
fermion at site~$l$. Later, we will set $J=1$ 
as our energy unit (bare bandwidth $W=4$).

The chain fermions are locally hybridized with reservoir fermions,
\begin{equation}
\label{Hyb}
\hat{S}=\sum_{l=1}^{L} 
\left(\gamma_{l,{\rm in}}\hat{c}_{l}^{\dagger}\hat{s}_{l}+
\gamma_{l,{\rm out}}\hat{s}_{l}^{\dagger}\hat{c}_{l}
\right) \; .
\end{equation}
The hybridization strength for
the fermion transfer from the reservoir to the chain,
$\gamma_{l,{\rm in}}$, can be different from
the transfer from the chain to the reservoir,
$\gamma_{l,{\rm out}}$.
In this work we focus on the Hermitian case,
$\gamma_{l,{\rm in}}=\gamma_{l,{\rm out}}$,
and the ratchet case,
$\gamma_{l,{\rm out}}=0$, where the chain fermions cannot return
to the reservoir sites.
We will study the translationally invariant case,
$\gamma_{l,{\rm in}}=\gamma$, and
the inhomogeneous case with
$\gamma_{l+1,{\rm in}}<\gamma_{l,{\rm in}}=J\exp(-l/L)$.

\begin{figure}[htb]
\begin{center}
\includegraphics[width=0.6\textwidth]{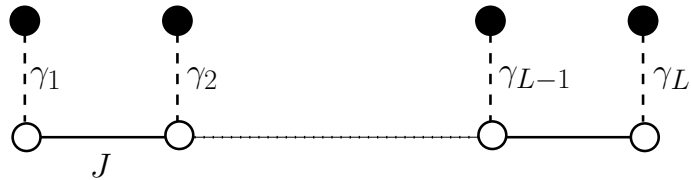}
\caption{Model for the particle injection into a chain. The solid circles represent the source sites, the open circles the chain sites. The solid and dashed lines denote the bonds with hopping amplitudes $J$
and $\gamma_l$ respectively.
  \label{fig:hamilt}}
\end{center}
\end{figure}

In Sect.~\ref{sec:interacting} we consider numerically the time evolution in
the presence of density-density interactions between the
fermions,
%\begin{eqnarray}
%\label{H_int}
%\hat{V}&=& V_{\rm c} \sum_{l=1}^{L-1} 
%\left(\hat{c}_{l}^{\dagger}\hat{c}_{l}-\frac{1}{2}\right)
%\left(\hat{c}_{l+1}^{\dagger}\hat{c}_{l+1}-\frac{1}{2}\right)
%+
%V_{\rm s} \sum_{l=1}^{L-1} 
%\left(\hat{s}_{l}^{\dagger}\hat{s}_{l}-\frac{1}{2}\right)
%\left(\hat{s}_{l+1}^{\dagger}\hat{s}_{l+1}-\frac{1}{2}\right)
%\nonumber \\
%&&+
%V_{\rm cs} \sum_{l=1}^{L} 
%\left(\hat{c}_{l}^{\dagger}\hat{c}_{l}-\frac{1}{2}\right)
%\left(\hat{s}_{l}^{\dagger}\hat{s}_{l}-\frac{1}{2}\right)
%\; .
%\end{eqnarray}
\begin{equation}
\label{H_int}
\hat{V}= \sum_{l} V_{\rm c} 
\bigl(\hat{c}_{l}^{\dagger}\hat{c}_{l}-\frac{1}{2}\bigr)
\bigl(\hat{c}_{l+1}^{\dagger}\hat{c}_{l+1}-\frac{1}{2}\bigr)
+ V_{\rm s} \bigl(\hat{s}_{l}^{\dagger}\hat{s}_{l}-\frac{1}{2}\bigr)
\bigl(\hat{s}_{l+1}^{\dagger}\hat{s}_{l+1}-\frac{1}{2}\bigr)
+V_{\rm cs} 
\bigl(\hat{c}_{l}^{\dagger}\hat{c}_{l}-\frac{1}{2}\bigr)
\bigl(\hat{s}_{l}^{\dagger}\hat{s}_{l}-\frac{1}{2}\bigr)
\; .
\end{equation}
All interaction parameters are chosen to be positive.
The total Hamiltonian is given by
\begin{equation}
\hat{H}=\hat{T}+\hat{S}+\hat{V} \; .
\end{equation}
The model is shown pictorially in Fig.~\ref{fig:hamilt}.

\subsection{Initial state and measured quantities}

In order to study the particle injection from the reservoir 
into the chain, we choose our initial state as
\begin{equation}
|\psi(t=0)\rangle=
|\psi_0\rangle = \prod_{l=1}^{L} \hat{s}_l^{\dagger} |{\rm vac}\rangle\; .
\end{equation}
At time $t=0$ all source sites are occupied and the chain is empty.

We are interested in the average particle density in the chain, $n_{c}(t)$,
the single-particle density matrix of the chain fermions,
$A_{p,k}(t)$, and the current density, $j_{c}(t)$.
These quantities are defined by
\begin{eqnarray}
A_{p,k}(t)&=&\langle \hat{a}_p^{\dagger}\hat{a}_k\rangle(t) \quad , \quad
\hat{a}_k=\sqrt{\frac{2}{L+1}}\sum_{l=1}^L 
\sin\left(\frac{\pi k l}{L+1}\right)\hat{c}_{l}
\quad (k=1,2,\ldots,L)\; ,\nonumber \\
n_{c}(t)&=&\frac{1}{L} \sum_{k=1}^{L}A_{k,k}(t)\; ,
\label{eq:defquantities}\\
j_{c}(t)&=&\frac{\rmi}{L}\sum_{l=1}^{L-1} 
\langle 
\hat{c}_{l+1}^{\dagger}\hat{c}_{l}-\hat{c}_{l}^{\dagger}\hat{c}_{l+1}
\rangle(t)
= 
-\frac{2\rmi}{L+1}\sum_{\substack{p,k=1\\ p\neq k}}^{L}
j(k,p)A_{p,k}(t) \; ,
\nonumber\\
j(k,p)&=&\frac{(1-(-1)^{p-k})}{L}
\frac{
\sin\bigl(p\pi/({L+1})\bigr)\sin\bigl(k\pi/({L+1})\bigr)
}{
\cos\bigl(k\pi/({L+1})\bigr)-\cos\bigl(p\pi/({L+1})\bigr)
} \nonumber \; .
\end{eqnarray}
For a homogeneous coupling, $\gamma_l=\gamma$, the single-particle density
matrix is diagonal, $A_{p,k}(t)=\delta_{k,p}A_{k,k}$, and the current density
is zero for all times, $j_{c}(t)=0$.

The long-time mean of a time-dependent quantity $X(t)$ is defined as 
\begin{equation}
\overline{X(t)}=\lim_{\tau\to\infty}\frac{1}{\tau}\int_0^{\tau}
X(t){\rm d}t \; .
\end{equation}

\section{Non-interacting fermions}
\label{sec:noninterating}

First, we address the question of how effective the injection of 
non-interacting fermions into the empty chain is. To this end, we
consider the average particle number, the single-particle density matrix,
and the current density in the chain, as functions of time. 

At short times we expect a linear increase of the particle density in
the chain because we start with all particles at the source sites.  At
later times, we shall see an oscillatory behavior of the particle
density because of the coherent quantum-mechanical movement of the
particles within and into the chain (and back onto the source sites
for the Hermitian case).  In the non-interacting case, a constant
average occupation is reached only for long times because the
two-level systems stay coherent for relatively long times so that the
oscillations in the particle density die out only slowly.
Interactions dampen these oscillations strongly and also modify the
value in the long-time limit considerably, see Sect.~\ref{sec:interacting}.

\subsection{Homogeneous chain filling}
\label{subsec:homchain}

We begin with the translationally invariant case 
for non-interacting particles,
$\gamma_{l,{\rm in}}=\gamma_{\rm in}$ and $\gamma_{l,{\rm out}}=\gamma_{\rm out}$
in eq.~(\ref{Hyb}).
For a homogeneous chain filling, the single-particle density matrix
is diagonal and the current density is zero for all times.

\subsubsection{Hermitian time evolution}
\label{subsub:hom-Hermitian_coupling}

In the Hermitian time evolution with 
$\gamma_{\rm in}=\gamma_{\rm out}=\gamma$ the model reads
\begin{equation}
\hh=-J\sum_{j=1}^{L-1}[\hc^\dagger_{j+1}\hc_{j}
+\hc^\dagger_{j+1}\hc_{j}]
+\gamma\sum_{j=1}^{L}[\hs^\dagger_j\hc_j+\hc^\dagger_j\hs_j]
=\sum_k[\varepsilon_k\ha^\dagger_k\ha_k
+\gamma(\ha^\dagger_k\hb_k+\hb^\dagger_k\ha_k)]\; ,
\label{freeh}
\end{equation}
where $\hat{b}_k$ denotes the Fourier transform of the operators $\hat{s}_l$
defined analogously to the Fourier transform of the chain operators in
eq.~(\ref{eq:defquantities}). 
The dispersion is given by $\varepsilon_k=-2J\cos k$ with $k=\pi n/(L+1)$.

Heisenberg's equations of motion provide the simplest way 
to solve this problem analytically.
The resulting coupled differential equations read
\begin{eqnarray}
\dot{\ha}_k&=&\rmi[\hh,\ha_k]=-\rmi(\varepsilon_k\ha_k+\gamma\hb_k)
\quad \textrm{and}\nonumber\\
\dot{\hb}_k&=&\rmi[\hh,\hb_k]=-\rmi\gamma\ha_k \; .
\end{eqnarray}
It is straightforward to write down a formal solution,
\begin{eqnarray}
\begin{pmatrix}\ha_k(t)\\ \hb_k(t)\end{pmatrix}
=e^{-\rmi \mathbf{\widetilde{M}_k}t}\begin{pmatrix}\ha_k\\ \hb_k\end{pmatrix}\; ,
\end{eqnarray} 
where
\begin{eqnarray}
\mathbf{\widetilde{M}_k}=
\begin{pmatrix}\varepsilon_k&\gamma\\\gamma&0\end{pmatrix}
\; .
\end{eqnarray}
Re-expressing the exponential matrix leads to
\begin{eqnarray}
\begin{pmatrix}\ha_k(t)\\ \hb_k(t)\end{pmatrix}=
e^{-\rmi\varepsilon_k t/2}\mathbf{M_k}(t)\begin{pmatrix}\ha_k\\ \hb_k\end{pmatrix}
\end{eqnarray}
with 
\begin{eqnarray}
M_{k,11}=M_{k,22}^*&=&
\cos\left(E_k t/2\right)
-\rmi\frac{\varepsilon_k \sin\left(E_k t/2\right)}{E_k}\; ,\nonumber\\
M_{k,12}=M_{k,21}&=&-\rmi \frac{2\gamma}{E_k}\sin\left(E_k t/2\right) \; ,\\
E_k&=&\sqrt{\varepsilon_k^2+4\gamma^2}\; . \nonumber
\end{eqnarray}
The zero-time correlation functions are very simple,
\begin{equation}
\langle\psi_0|\ha^\dagger_k\ha_k|\psi_0\rangle
=
\langle\psi_0|\ha^\dagger_k\hb_k|\psi_0\rangle=0\quad  ,\quad
\langle\psi_0|\hb^\dagger_k\hb_k|\psi_0\rangle=1\; .
\end{equation}
The single-particle density matrix is diagonal and given by
\begin{equation}
\label{Hermitian_nt1}
A_{k,k}(t)=\left|M_{k,12}\right|^2
=\frac{4\gamma^2}{E_k^2}\sin^2\left(E_k t/2\right) \; .
\end{equation}
It describes the beating of particles in the two-level systems
with energies $\varepsilon_{k;1,2}=(1/2)(\varepsilon_k\pm E_k)$ which
result from the hybridization of the bath particles at zero energy
with the chain particles with dispersion $\varepsilon_k$.

The average particle number in the chain becomes
\begin{equation}
\label{Hermitian_nt2}
n_c(t)=\frac{1}{L}\sum_k
\frac{4\gamma^2}{E_k^2}\sin^2\left(E_k t/2\right)  \; .
\end{equation}
Its long-time mean is straightforwardly extracted,
\begin{eqnarray}
\overline{n_c(t)}=\frac{1}{2L}\sum_k 
\frac{4\gamma^2}{E_k^2} \; .
\end{eqnarray}
In principle, the oscillations around
the mean will not die out for long times. For
any finite chain there exists a recurrence
time $t^{\epsilon}_{\rm rec}\neq 0$ such that for
the numerator of the sum~(\ref{Hermitian_nt2}) the relation
$\sin^2(t^{\epsilon}_{\rm  rec}E_k/2)<\epsilon/L$ holds for all allowed 
$k$-values. However, this Poincar\'e time is very large even for
systems $L=50$ because the energies $E_k$ are not commensurate.

In the thermodynamic limit ($L\to\infty$)
eq.~(\ref{Hermitian_nt2}) reduces to
\begin{equation}
\label{cnt}
n_c(t)=\frac{2}{\pi}\int_{0}^{J/\gamma}
\frac{{\rm d}\varepsilon}{\sqrt{(J/\gamma)^2-\varepsilon^2}}
\frac{\sin^2(\gamma t\sqrt{1+\varepsilon^2})}{1+\varepsilon^2}\; .
\end{equation}
In this case the oscillations in $n_c(t)$ do vanish for $t\to\infty$
because the infinitely many two-level systems dephase
completely. We find
\begin{eqnarray}
\label{cnt2}
n_c(t\to\infty)= \frac{1}{2}\frac{1}{\sqrt{1+(J/\gamma)^2}}\; .
\end{eqnarray}
For $J/\gamma\to 0$ only half of the chain sites are occupied on average.
Recall that, for $J=0$, 
we describe a collection of independent two-level systems 
where each particle oscillates between two levels such that it is 
equally likely to be found on the chain and on the bath site.

For $J/\gamma\ll 1$, the integral~(\ref{cnt}) 
can be solved to leading order,
\begin{equation}
\label{Asympt_decay}
n_c(t)\approx\frac{1}{2} 
- \frac{1}{2}J_0(tJ^2/(2\gamma))
\left[\cos(2\gamma t)\cos(tJ^2/(2\gamma))
+\sin(2\gamma t)\sin(tJ^2/(2\gamma))\right]
\; .
\end{equation}
Here, the trigonometric functions describe the fast oscillations around
the mean value whereas the Bessel function $J_0(tJ^2/2\gamma)$ gives the
envelope. The oscillations around the mean value
decay in the long time limit proportional to 
$\sqrt{2\gamma/J^2t}$. The slow oscillatory decay proportional to $1/\sqrt{t}$
results from the square-root divergence in the single-particle
density of states in one dimension.

\begin{figure}[htb]
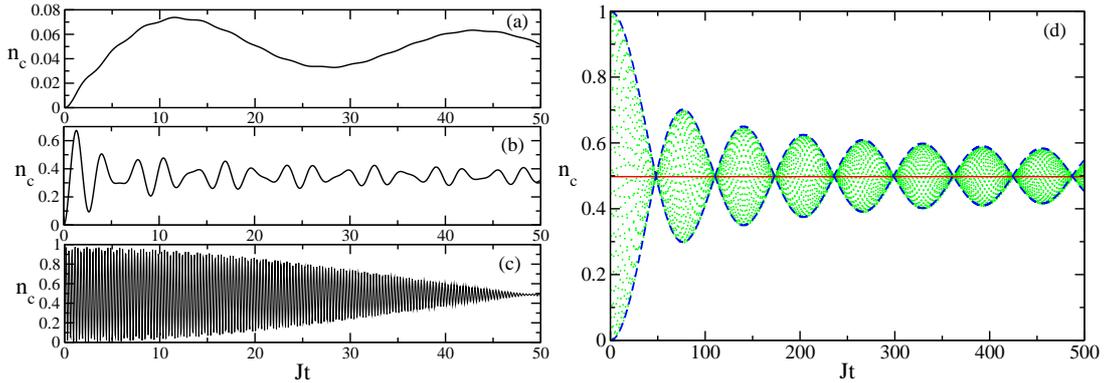

\begin{center}
\includegraphics*[width=0.48\textwidth]{L50_hermitian_various_gamma.eps}
\includegraphics*[width=0.48\textwidth]{ContLimit_hermitian_gamma10.eps}
\end{center}
\caption{Particle density in the chain $n_c(t)$ 
for non-interacting fermions (Hermitian time evolution). 
Left: a chain with $L=50$ sites and
(a)~$\gamma/J=0.1$, (b)~$\gamma/J=1$, (c)~$\gamma/J=10$;
(d): thermodynamic limit for $\gamma/J=10$
(circles: time steps $\delta t=0.1$). 
The solid line is the long-time limit, the dashed lines show the envelope
function according to eq.~(\protect\ref{Asympt_decay}).
\label{Fig1}}
\end{figure}

In Fig.~\ref{Fig1} we show results for
different ratios of $\gamma/J$ for a chain with $L=50$ sites
for times $Jt\leq 50$.
For a finite chain, the average particle number will never become
constant. Instead, revival oscillations will occur because the Hilbert
space has a finite dimension.
In the thermodynamic limit, the oscillations 
slowly decay to a constant average particle number in the
chain given by eq.~(\ref{cnt}), see Fig.~\ref{Fig1}(d), plotted for $\gamma/J=10$ and times $Jt\leq 500$.

\begin{figure}[htbp]
\begin{center}
\includegraphics*[width=0.9\textwidth]{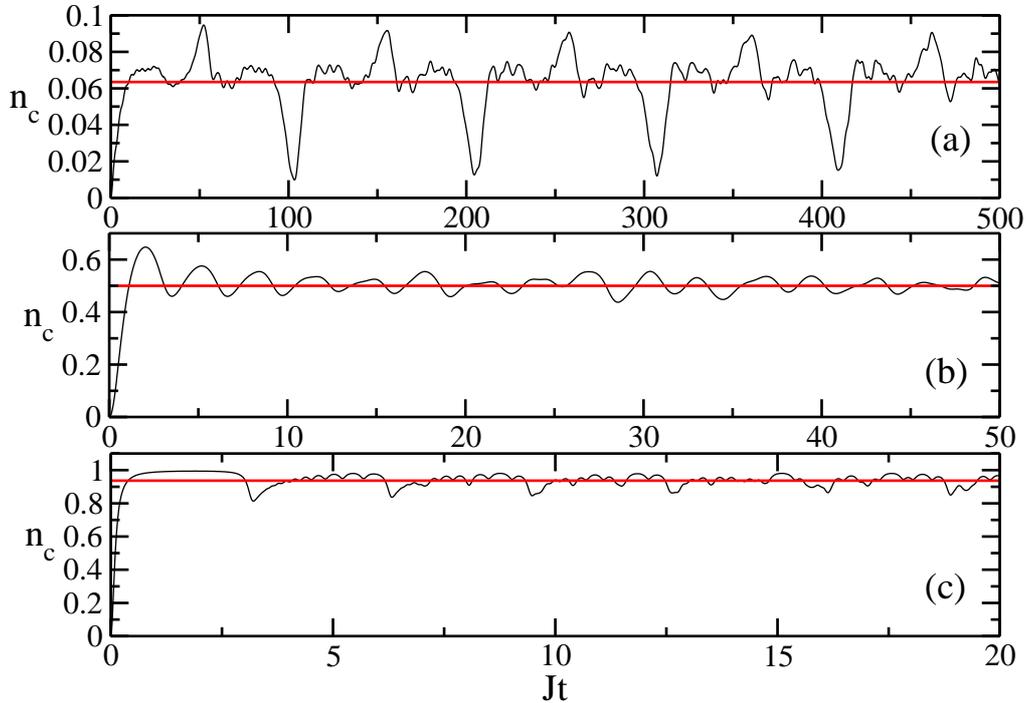}
\end{center}
\caption{Particle density in the chain $n_c(t)$ for non-interacting
  fermions on a chain with $L=50$ sites for (a)~$\gamma_{\rm
    in}/J=0.1$, (b)~$\gamma_{\rm in}/J=1$, and (c)~$\gamma_{\rm
    in}/J=10$ (ratchet time evolution).  The straight lines denote the
  mean values.  Note the different scales on the axis.\label{Fig3}}
\end{figure}

As a comparison, we show results for the ratchet case in
Fig.~\ref{Fig3} for the same values of $\gamma_{\rm in}=\gamma$ 
($\gamma_{\rm out}=0$) and different time spans.
Even in the ratchet case, not all particles are injected into the
chain at long times. Only for $\gamma_{\rm in}\to \infty$ does the
particle number in the chain approach unity, see
eq.~(\ref{Non_Herm_long_time}). 
In addition, one still observes 
fluctuations so that the particle density
in the chain is not a monotonically increasing function. 
The motion of the fermions on the chain
leads to Pauli blocking and to quantum-mechanical interference effects
which result in the observed incomplete and oscillatory chain filling.
In comparison with the Hermitian time evolution 
it can be seen that in the ratchet case
the quantum-mechanical fluctuations 
are smaller and the chain filling is bigger for the same parameter~$\gamma/J$.

\begin{figure}[htbp]
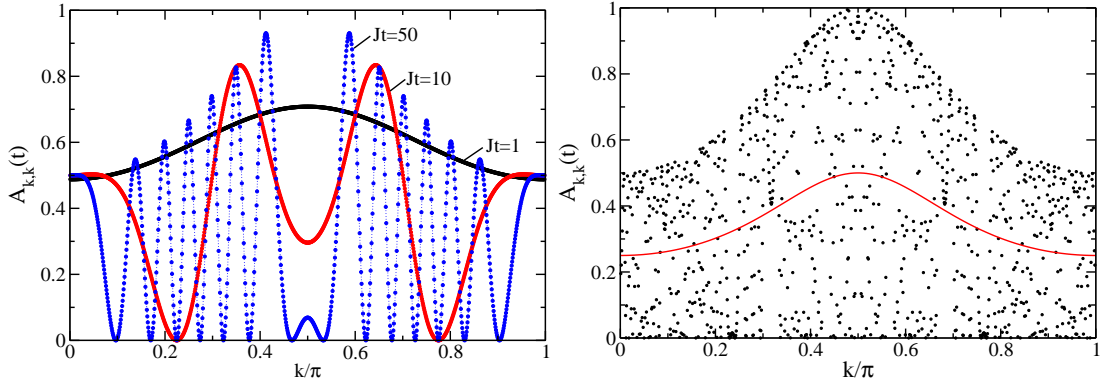

\begin{center}
\includegraphics*[width=0.48\textwidth]{L1000_gamma1_Akk_various_t.eps}
\includegraphics*[width=0.48\textwidth]{L1000_gamma1_Akk_t5000.eps}
\end{center}
\caption{Density matrix $A_{k,k}(t)$ in the Hermitian time evolution with
$\gamma/J=1$ for a chain of length $L=1000$ as a function of the
discrete momenta $k=\pi n/(L+1)$ for times $Jt=1,10,50$ (left)
and $Jt=5000$ (right) where the 
solid line is the long-time mean.\label{Fig7}}
\end{figure}

In Fig.~\ref{Fig7} we show the single-particle
density matrix $A_{k,k}$ for Hermitian time evolution with
$\gamma/J=1$ at various times.
At small times all particles enter the chain at roughly the same
moment which leads to a relatively flat distribution for $Jt=1$.
At larger times, the individual dynamics of each $k$-point becomes
visible and a lot of maxima and minima have developed already by $Jt=50$.
At very large times, $Jt=5000$, the map $A_{k,k}(t)$, corresponding to a large collection
of undamped oscillators, displays a non-monotonic
(``chaotic'') distribution.
The size of the fluctuations around the time-averaged distribution 
is as large as its mean. The system does not relax.

\begin{figure}[htbp]
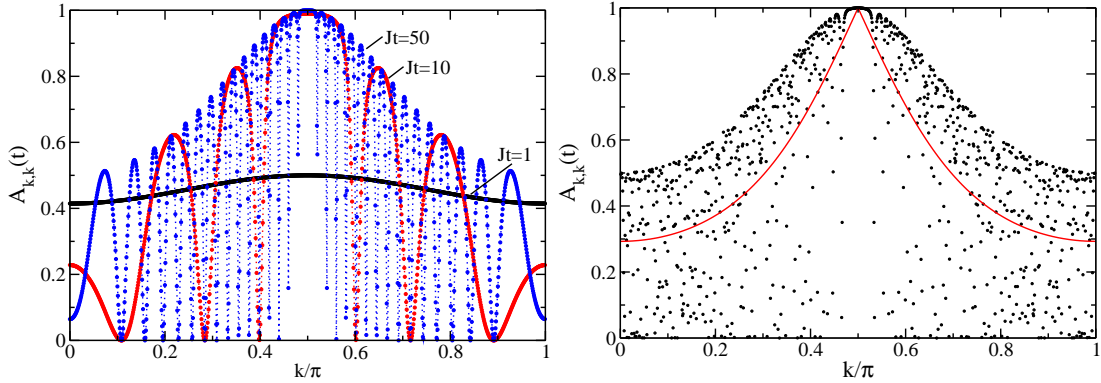

\begin{center}
\includegraphics*[width=0.48\textwidth]{L1000_nonhermitian_gamma1_Akk_various_t.eps}
\includegraphics*[width=0.48\textwidth]{L1000_gamma1_Akk_t5000_non_Herm.eps}
\end{center}
\caption{Density matrix $A_{k,k}(t)$ in the ratchet case with
$\gamma_{\rm in}/J=1$ for a chain of length $L=1000$ as a function of the
discrete momenta $k=\pi n/(L+1)$ for times $Jt=1,10,50$ (left) 
and $Jt=5000$ (right) where the 
solid line is the long-time mean.\label{Fig9}}
\end{figure}

The corresponding results for the single-particle density matrix $A_{k,k}$
for non-interacting electrons in the ratchet case 
are shown in Fig.~\ref{Fig9}. 
The average distribution is sharply peaked at
$k=\pi/2$, see eq.~(\ref{NonHermitian_Akk_ave}). This means that 
for large times the
$k$-points near $\pi/2$ show very little variation with time while the
$k$-points away from $\pi/2$ are still strongly oscillating in time.

\subsubsection{Ratchet case}
\label{subsub:inhom-non_Hermitian_coupling}

Here we provide the analytical expressions on which the data in 
Figs.~\ref{Fig3} and~\ref{Fig9} are based.
We treat the translationally invariant ratchet
case, $\gamma_{l,{\rm out}}=0$ and $\gamma_{l,{\rm in}}=\gamma$. The
Fourier transformed Hamiltonian reads
\begin{equation}
\hh=\sum_k[\varepsilon_k\ha^{\dagger}_k\ha_k+\gamma\ha^{\dagger}_k\hb_k]
\equiv \hh_0+\hat{S}\; ,
\end{equation}
where $\hh_0$ is the Hermitian and $\hat{S}$ is the non-Hermitian part. 
In the interaction picture representation we find
\begin{equation}
\hat{S}(t)=\gamma\sum_k e^{\rmi\varepsilon_k t}\ha^{\dagger}_k\hb_k\; .
\end{equation}
In this case one can explicitly calculate the $S$-matrix 
because $\hat{S}(t)$ commutes with itself at different times and therefore
one can directly integrate over times in $\hat{S}(t)$. 
The wave function becomes
\begin{equation}
|\psi(t)\rangle= \prod_{k=1}^L\left( 
1+ \gamma g_k(t)\hat{b}_k\hat{a}_k^{\dagger}\right)
|\psi(0)\rangle \quad , \quad 
g_k(t)=\frac{1-e^{-\rmi \varepsilon_k t}}{\varepsilon_k}\; .
\end{equation}
For the single-particle density matrix we find
\begin{equation}
\label{NonHermitian_Akk}
A_{k,k}(t)= \frac{\gamma^2|g_k(t)|^2}{1+\gamma^2|g_k(t)|^2}=
\frac{\sin^2(\varepsilon_kt/2)}
{\varepsilon_k^2/(4\gamma^2)+\sin^2(\varepsilon_k t/2)}\; .
\end{equation}
In contrast to the Hermitian time evolution, 
the bare energy levels of the bath and the
chain fermions do not hybridize. Therefore, the bare dispersion 
$\varepsilon_k$ appears as the beating frequency. Also note the time-dependence
of the denominator which is absent for the Hermitian time evolution, 
see eq.~(\ref{Hermitian_nt1}).

Again, we note that it is always possible to find a recurrence time
$t^{\epsilon}_{\rm rec}$ such that the average particle number in the
chain $n_c(t^{\epsilon}_{\rm rec})=\sum_k A_{k,k}(t^{\epsilon}_{\rm rec})/L<\epsilon$ 
for any given $\epsilon$.
At first, these recurrences seem counter-intuitive because, in the ratchet case,
the amplitude for particles going back from the chain onto the
source sites is zero. This clearly demonstrates the quantum character of
the problem.

The time averaged distribution function is given by
\begin{eqnarray}
\label{NonHermitian_Akk_ave}
\overline{A_{k,k}(t)}=1-\frac{|\epsilon_k|}{E_k}\; ,
\end{eqnarray}
which is sharply peaked at $k=\pi/2$.
In the thermodynamic limit we therefore find that 
the density of particles in the chain is given by
\begin{equation}
\label{Non_Herm_long_time}
n_c(t\to\infty) =1-\frac{2}{\pi}\arctan\left(\frac{J}{\gamma}\right) \; .
\end{equation}
For a collection of independent two-level systems of bath and chain sites
($J=0$), the chain is completely filled for large times because
the particles cannot return to the bath sites.
For $J/\gamma\neq 0$, however, 
we have $n_c(t\to\infty)<1$, i.e., the chain is not completely filled
despite the ratchet condition.
Due to the motion of the particles on the chain,
some bath particles are prevented from falling down into the chain
because their chain sites are blocked by other particles 
(Pauli exclusion principle).

\subsection{Inhomogeneous chain filling}
\label{subsub:inhom-Hermitian_coupling}

For an inhomogeneous coupling, the problem 
must be treated numerically. It involves the diagonalization of
the Hamiltonian (Hermitian time evolution) or the inversion of the
self-energy matrix (ratchet case).

\subsubsection{Hermitian time evolution}

We rewrite the Hamilton operator as 
\begin{equation}
\hat{H}= \sum_{i,j=1}^{2L}H_{i,j}\hat{p}_i^{\dagger}\hat{p}_j
\end{equation}
with $\hat{p}_{2m-1}=\hat{s}_{m}$, $\hat{p}_{2m}=\hat{c}_{m}$
for $m=1,2,\ldots,L$
and $H_{2m-1,2m}=H_{2m,2m-1}=\gamma_m$, $H_{2m+2,2m}=H_{2m,2m+2}=-J$
for $m=1,2,\ldots,L-1$.
The Hamiltonian is diagonal in a new single-particle basis,
\begin{equation}
\hat{H}=\sum_{N=1}^{2L}E_N\hat{\xi}_N^{\dagger}\hat{\xi}_N \quad , \quad
\hat{\xi}_N= \sum_{j=1}^{2L}U_{j,N}\hat{p}_j \quad, \quad
\hat{p}_j= \sum_{N=1}^{2L}U^*_{j,N}\hat{\xi}_N=\sum_{N=1}^{2L}U_{j,N}\hat{\xi}_N
\; ,
\end{equation}
where we used in the last step that the transformation matrix~$U$, which is determined numerically, is real.
The single-particle density matrix in momentum space is
best calculated from the corresponding expression in position space,
\begin{equation}
A_{p,q}(t)=\sum_{l,r}\frac{2}{L+1}\sin\left(\frac{\pi p l}{L+1}\right)
\sin\left(\frac{\pi q  r}{L+1}\right) C_{l,r}(t)
\end{equation}
with 
\begin{eqnarray}
C_{l,r}(t) &=& \langle \psi(t) | \hat{p}_{2l}^{\dagger}\hat{p}_{2r}
|\psi(t)\rangle
\nonumber \\
&=& \sum_{N,M=1}^{2L} e^{\rmi(E_N-E_M)t}
\left( \sum_{m=1}^{L}U_{2m-1,N}U_{2m-1,M}\right) U_{2l,N}U_{2r,M}\;.
\label{Akp-nonsymmetric-hermitian}
\end{eqnarray}
Here we used the time evolution~(\ref{eq:schroedinger}) and
the fact that all bath sites are occupied at time $t=0$ so that 
$\langle \psi(0) | 
\hat{p}_{2m-1}^{\dagger}\hat{p}_{2n-1}
|\psi(0)\rangle=\delta_{m,n}$.
The calculation of $A_{p,q}(t)$ involves five lattice summations
which are performed numerically.

\begin{figure}[htbp]
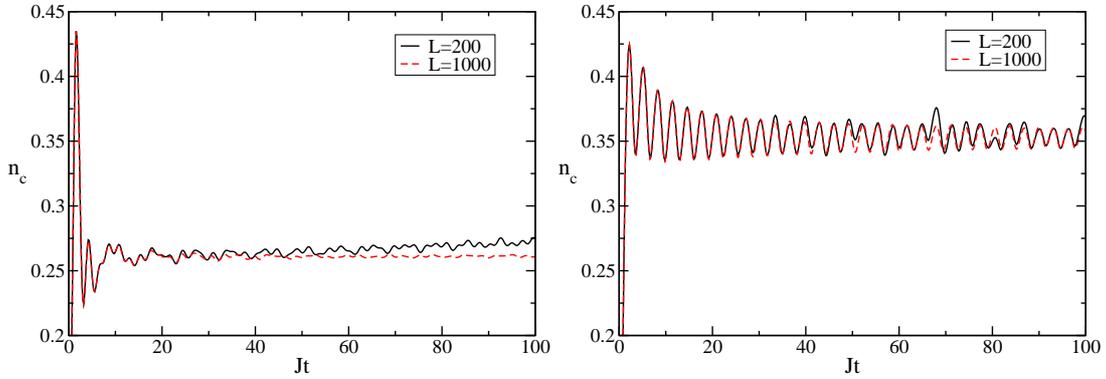

\begin{center}
%\begin{tabular}{@{}ll@{}}
\includegraphics*[width=0.48\textwidth]{unit_nonsym.eps}
\includegraphics*[width=0.48\textwidth]{non_nonsym.eps}
%\end{tabular}
\end{center}
\caption{Particle density in the chain $n_c(t)$ 
for non-interacting fermions with hopping $\gamma_l=J\exp(-l/L)$ on 
a chain with $L=200$ ($L=1000$) sites for unitary time evolution (left)
and ratchet time evolution (right).
\label{fig:nonsymm-nonint-nc}}
\end{figure}

For an inhomogeneous coupling, $\gamma_l=J\exp(-l/L)$, the
particle-density in the chain $n_c(t)$ is qualitatively similar to the
case of a symmetric filling for $\gamma=J$.  In
Fig.~\ref{fig:nonsymm-nonint-nc}, we show the filling of the chain for
the unitary time evolution and the ratchet case for times $Jt\leq 100$.

In the unitary case, the fluctuations in the particle density are much
smaller for the inhomogeneous hybridizations than for constant
hybridizations.  In addition, larger system sizes reduce the
fluctuations further as can be seen from the comparison for system
sizes $L=200$ and $L=1000$.  In any case the particle density reaches
its long-time value for times $Jt\gtrsim 1000$ with small fluctuations
around it. These reflect the transient current density in the system,
see Fig.~\ref{fig:non-interacting-currents}. In the ratchet case, we
observe that the particle density in the long-time limit for the same
set of parameters is higher than in the unitary case.  Furthermore,
the oscillations around this long-time mean are more regular and decay
algebraically.

\begin{figure}[htbp]
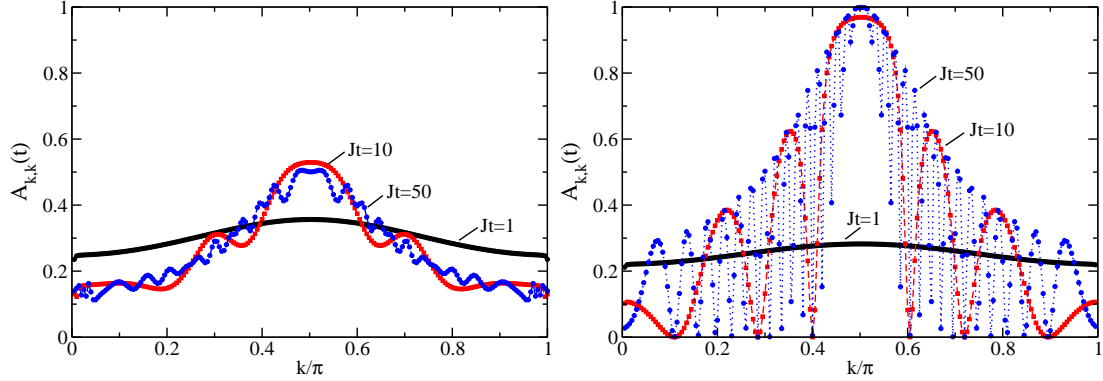

\begin{center}
\includegraphics*[width=0.48\textwidth]{diag_nonsym_unit.eps}
\includegraphics*[width=0.48\textwidth]{diag_nonsym_nonunit.eps}
\end{center}
\caption{Diagonal part of the single-particle density matrix $A_{k,k}(t)$
for an inhomogeneous chain hopping, $\gamma_l=J\exp(-l/L)$, with $L=200$,
for the Hermitian time evolution (left) and the ratchet case (right).
\label{fig:akk-beni}}
\end{figure}

In the case of an inhomogeneous coupling, $\gamma_l=J\exp(-l/L)$, the
single-particle density matrix is no longer diagonal. The solution of
the equations~(\ref{Akp-nonsymmetric-hermitian})
and~(\ref{eq:Sigmasolution}) for $L=200$ sites shows that dephasing is
very efficient for the non-diagonal elements $A_{k,p\neq k}$.  Even at
short times, $Jt\sim 1$, the single-particle density matrix is almost
diagonal with $A_{k,p\neq k}(t)$ being an order of magnitude smaller
than $A_{k,k}(t)$ for all $k,p$. Further decay in time of the
off-diagonal elements is only algebraic thus leading to an
algebraically decaying current density, see the following discussion.
In Fig.~\ref{fig:akk-beni} we show the time evolution of the diagonal
components of the single-particle density matrix, $A_{k,k}(t)$. The
qualitative behavior of $A_{k,k}(t)$ is the same for the homogenous
and the inhomogeneous coupling. Note that the different chain fillings
for the same parameter set lead to a different overall scale for
$A_{k,k}$.  The comparison shows that the frequency and the size of
the beating oscillations is smaller in the Hermitian time evolution
than in the ratchet case.  For an inhomogeneous hybridization, $k$~is
no longer a good quantum number.  Apparently, the mode coupling in the
Hermitian case noticeably reduces the fluctuations of $A_{k,k}(t)$
around its time average.  In the ratchet case, the Hamiltonian can
still be written as a sum over commuting $k$-modes, see
eq.~(\ref{eq:nonsymmnonhermitianH}).  Therefore, we observe a
pronounced beating in $A_{k,k}(t)$ even for inhomogeneous
hybridizations.  As we shall see in Sect.~\ref{sec:interacting},
interactions introduce a strong scattering between particles so that
oscillations of the $k$~modes are absent and $A_{k,k}(t)$ quickly
relaxes to its steady-state distribution.

The non-diagonal components of $A_{k,p}$ enter the 
expression for the current density~(\ref{eq:defquantities}).
A small transient current sets in at $t>0$
and decays to zero for long times $Jt\gg 1$.
For our case study we choose $\gamma_l=J\exp(-l/L)$ 
so that we insert particles more readily
in the left part of the chain than in the right. Originally, there 
is a current from the left to the right of the chain ($j_c>0$)
which is reflected at the right chain end
so that $j_c(t)$ changes sign as a function of time.
As for the particle density,
the current density shows oscillations which decay algebraically. 

In Fig.~\ref{fig:non-interacting-currents} we show the 
transient current density for the Hermitian time evolution and the ratchet case
for $L=200$ sites. The amplitudes differ for
the same parameter set which is mostly due to the fact that 
there are more particles in the chain for the ratchet case.
The frequency of the current oscillations is comparable.
The current oscillations correspond to 
a swapping of the particles between the right and left chain boundaries.
Since the mobile particles are concentrated around $\varepsilon(k_0)=0$
($k_0=\pi/2)$,
their dominant velocity is given by 
$v_0=\left.(\partial \varepsilon(k))/(\partial k)\right|_{k=k_0}=2$. 
Therefore, their travel time between
the chain boundaries is $T/2\approx L/v_0= L/2$ so that the oscillation period 
is $T\approx L$. As seen in Fig.~\ref{fig:non-interacting-currents},
this rule applies better for the ratchet case where the distribution 
$A_{k,p}(t)$ is more peaked around $(k,p)=(\pi/2,\pi/2)$.
\begin{figure}[htbp]
\begin{center}
\includegraphics*[width=0.46\textwidth]{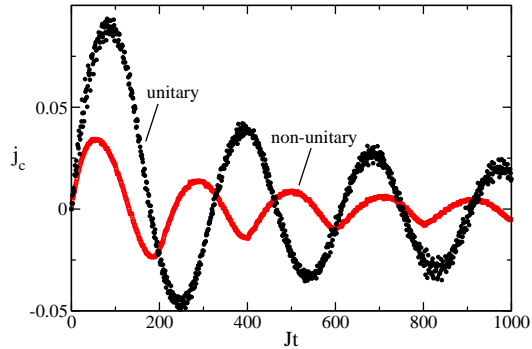}\\
\end{center}
\caption{Current density $j_c(t)$ for non-interacting electrons for
  $\gamma_l=J\exp(-l/L)$, unitary and non-unitary
  time evolution, for $L=200$ sites.
\label{fig:non-interacting-currents}}
\end{figure}

\subsubsection{Ratchet case}
\label{subsub:inhom-non_Hermitian-coupling}

Here we derive the formulae on which the data in 
Figs.~\ref{fig:nonsymm-nonint-nc}, \ref{fig:akk-beni} 
and~\ref{fig:non-interacting-currents} are based.
Even in the case of a non-symmetric coupling we may write the
ratchet Hamiltonian in the form
\begin{equation}
\hat{H}=\sum_{k=1}^L \hat{H}_k\quad , \quad 
\hat{H}_k=\varepsilon_k\hat{a}_k^{\dagger}\hat{a}_k
+\hat{a}_k^{\dagger}\hat{B}_k
\label{eq:nonsymmnonhermitianH}
\end{equation}
with $\left[ \hat{H}_k,\hat{H}_p\right]_{-}=0$ for all $k,p$.
In eq.~(\ref{eq:nonsymmnonhermitianH}) we introduced the pseudo-fermionic
operators
\begin{equation}
\hat{B}_k=\sqrt{\frac{2}{L+1}} \sum_{l=1}^L \gamma_l 
\sin\left(\frac{\pi k l}{L+1}\right) \hat{s}_l \;, 
\end{equation}
which obey $\left\{ \hat{B}_k,\hat{B}_p\right\}_{+}=
\left\{ \hat{B}_k^{\dagger},\hat{B}_p^{\dagger}\right\}_{+}=0$ for all $k,p$.
In particular, $(\hat{B}_k)^2=(\hat{B}_k^{\dagger})^2=0$.
However, the operators and their adjoint do not anti-commute,
\begin{equation}
\left\{ \hat{B}_k,\hat{B}_p^{\dagger}\right\}_{+}=
\frac{2}{L+1} \sum_{l=1}^L |\gamma_l|^2
\sin\left(\frac{\pi k l}{L+1}\right) 
\sin\left(\frac{\pi p l}{L+1}\right) 
\equiv H(k,p) \; ,
\label{eq:defHPQ}
\end{equation}
where we introduced the commutator function $H(k,p)$.
For $\gamma_l=\gamma$, we have $\hat{B}_k=\gamma\hat{b}_k$
and $H(k,p)=|\gamma|^2\delta_{k,p}$.

The wave function can be obtained 
as in Sect.~\ref{subsub:inhom-Hermitian_coupling},
\begin{equation}
|\psi(t)\rangle= \prod_{k=1}^L\left( 
1+ g_k(t)\hat{B}_k\hat{a}_k^{\dagger}\right)
|\psi(0)\rangle \quad , \quad 
g_k(t)=\frac{1-e^{-\rmi \varepsilon_k t}}{\varepsilon_k}
\label{eq:psi-t-inhomratchet}
\end{equation}
because the components $\hat{H}_k$ of $\hat{H}$ commute with each other.
As shown in appendix~\ref{app:D}, the single-particle density matrix
can be written as
\begin{equation}
A_{p,q}=g_qg_p^*\left[H(p,q)+\sum_{k_1,k_2=1}^L
H(p,k_1)\Sigma(k_1,k_2)H(k_2,q)\right] \; ,
\label{eq:solveApq}
\end{equation}
where $\Sigma(k_1,k_2)$ are the entries of the self-energy matrix 
$\mathbf{\Sigma}$.
This is obtained from the solution of the matrix equation
\begin{equation}
\mathbf{\Sigma}=-(\mathbf{1}+ \mathbf{f}\cdot \mathbf{H})^{-1}\mathbf{f} 
\; .
\label{eq:Sigmasolution}
\end{equation}
Here, the entries of the matrices $\mathbf{1}$, $\mathbf{H}$ and 
$\mathbf{f}$ are $1_{a,b}=\delta_{a,b}$ (unit matrix),
$H_{a,b}=H(a,b)$ (commutator matrix) and $f_{a,b}=\delta_{a,b}f_a$, $f_a=|g_a|^2$
(diagonal vertex matrix).
The numerical solution requires the inversion of an $L\times L$ matrix.

\section{Interacting fermions}
\label{sec:interacting}

In the thermodynamic limit and for non-interacting electrons,
the average particle number in the chain
goes to a constant at long-times only due
to decoherence. On the other hand, 
a true relaxation of the particle density and other physical
quantities can only take place when interactions are included. 

To investigate the model with
the density-density interaction terms, eq.~(\ref{H_int}), included, we
use an adaptive time-dependent density matrix renormalization group
($t$-DMRG) algorithm. In DMRG the wave function of the problem under
consideration is approximated by a matrix-product state
in a truncated Hilbert space. The time
evolution of this state is then calculated by using a second-order
Trotter-Suzuki decomposition of the time-evolution operator. For
details of the algorithm the reader is referred to
Refs.~\cite{DaleyKollath,WhiteFeiguin}. In our calculations we use a
Trotter slicing $\delta t=0.1$ and keep $\chi=250$ states.
It is known that the matrix dimension necessary to represent 
faithfully the
time-evolved state increases exponentially with time. The $t$-DMRG
algorithm is therefore only capable of describing the time evolution
at short and intermediate times. 

\subsection{Particle number in the chain}
\label{sub:n_interacting}

\begin{figure}[htbp]
\begin{center}
\includegraphics*[width=0.7\textwidth]{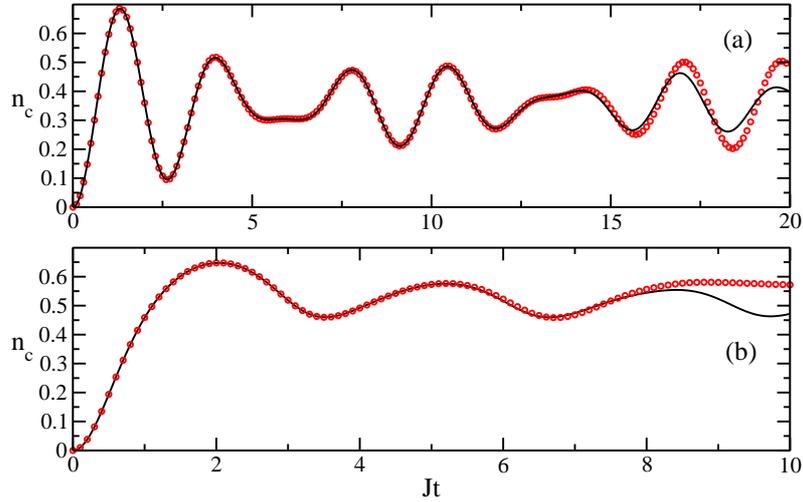}
\end{center}
\caption{Particle density in the chain $n_c(t)$ 
for non-interacting fermions on 
a chain with $L=50$ sites (a) for the Hermitian time evolution
($\gamma/J=1$) and (b) for the ratchet time evolution ($\gamma_{\rm in}/J=1$).
Full lines: analytical result; dots: data from $t$-DMRG 
($\chi=250$ states kept).\label{Fig4}}
\end{figure}

In order to test the algorithm and the limitations for the time evolution,
we perform numerical calculations for the particle density
of non-interacting fermions with a homogeneous
coupling where analytical results are available, 
see Sect.~\ref{subsec:homchain}.
In Fig.~\ref{Fig4} we compare numerical results obtained 
with the $t$-DMRG algorithm with exact
results in the non-interacting case, both for the unitary and 
ratchet time evolution. As seen from the figure,
the $t$-DMRG results ($\chi=250$ states kept)
are reliable for times $Jt\sim 14$ ($Jt\sim 8$)
in the Hermitian (ratchet) time evolution.

Apparently, the $t$-DMRG is not able to reach the limit of large times when
the two-level systems will have dephased completely. 
However, the algebraic decay of the phase coherence in $n_{c}(t)$
as given in eq.~(\ref{Asympt_decay}) and 
as shown in Fig.~\ref{Fig1} is special to non-interacting fermions.
Interactions between the particles lead to a much faster
dephasing and, moreover, to a true relaxation.
\begin{figure}[htbp]
\begin{center}
\includegraphics*[width=0.7\textwidth]{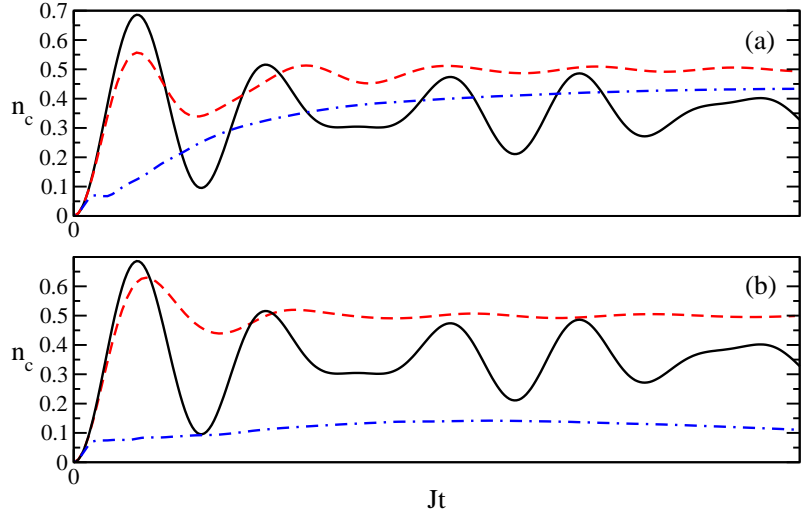}
\end{center}
\caption{Particle density in the chain $n_c(t)$ 
for $L=50$ lattice sites and $\gamma/J=1$
for the Hermitian time evolution.
We show $t$-DMRG results for non-interacting fermions 
($V_c=V_s=V_{cs}=0$, solid lines) in comparison
with interacting fermions:
(a) $V_c=V_{cs}=2, V_s=0$ (dashed line) and $V_c=V_{cs}=8, V_s=0$
(dot-dashed line);
(b) $V_c=V_{cs}=0, V_s=2$ (dashed line) 
and $V_c=V_{cs}=0, V_s=8$ (dot-dashed line).\label{Fig5}}
\end{figure}
In Fig.~\ref{Fig5} we show examples for the particle density
in the chain for the Hermitian time evolution ($\gamma/J=1$)
in the presence of various types and strengths of the
density-density interaction. In all cases, 
the interactions damp the oscillations and the average
particle number reaches its limiting value much faster.

\begin{figure}[htbp]
\begin{center}
\includegraphics*[width=0.7\textwidth]{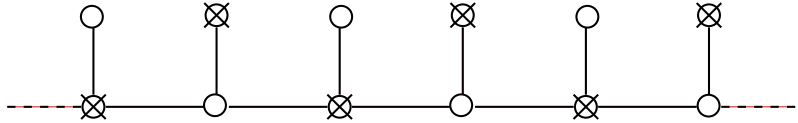}
\end{center}
\caption{Charge ordering in a Wigner crystal state. The crosses mark the occupied sites.\label{fig:Wigner}}
\end{figure}

Depending on the type of interactions present, the energy of the system in its initial state
\begin{equation}
\label{E_ini}
E_{\rm ini} = \langle \Psi_0| H |\Psi_0\rangle
\end{equation}
is different. Since the Hamiltonian is a conserved quantity, the
energy cannot change during time evolution. In the case where $V_s=0$
(Fig.~\ref{Fig5}a) the system starts in a state with $E_{\rm ini}=0$.
Another state with zero potential energy for this type of
density-density interactions is the twofold degenerate charge ordered
state where chain and source sites are occupied alternatingly (Wigner
lattice).  This is shown pictorially in Fig.~\ref{fig:Wigner}. At
least in the limit where $V_c=V_{cs}$ is large, i.e., where the
potential energy dominates over the kinetic energy, we therefore
expect the system to reach a stationary state which is close to this
charge ordered state. Therefore, the average particle density
approaches $n_c(Jt\gg 1)=1/2$ (see Fig.~\ref{Fig5}a) for large
interactions which is {\sl larger\/} than in the non-interacting case.
Apparently, the Coulomb interaction reduces the degree of
delocalization of electrons on the chain whereby the Pauli blocking
becomes less effective.  The reduction of the Pauli pressure increases
the density of fermions in the chain.

The case $V_c=V_{cs}=0$ with $V_s\neq 0$ (Fig.~\ref{Fig5}b) has an
energy $E_{\rm ini}= (L-1)V_s$. For the situation where $V_s$ is
smaller than the bandwidth, energy conservation can be fulfilled by
compensating the loss of potential energy by kinetic energy which the
fermions gain by entering the chain. Numerically we find that the
density of these almost free fermions in the chain can reach
$n_c(Jt\gg 1)\approx 1/2$, see Fig.~\ref{Fig5}b. For $V_c=V_{cs}=0$
with $V_s$ larger than the bandwidth, on the other hand, there is no
possibility of compensating the loss in potential energy with kinetic
energy in the chain. Hence the particles are almost all confined to
the bath sites by energy conservation and $n_c(Jt\gg 1)$ remains small, see Fig.~\ref{Fig5}b.

\begin{figure}[htbp]
\begin{center}
\includegraphics*[width=0.7\textwidth]{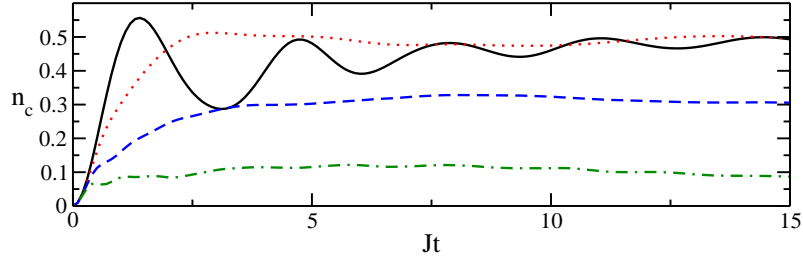}
\end{center}
\caption{Particle density in the chain $n_c(t)$ 
for $L=50$ lattice sites and $\gamma/J=1$
for the Hermitian time evolution.
We show $t$-DMRG results for interacting fermions: 
$V_c=V_s=V_{cs}=1$ (solid line),
$V_c=V_{cs}=V_s=2$ (dotted line), $V_c=V_{cs}=V_s=3$ (dashed line), and $V_c=V_{cs}=V_s=4$
(dot-dashed line).
\label{fig:densityallVs}}
\end{figure}

The fast damping of the oscillations of the particle density is a
genuine interaction effect which does not depend on the formation of a
charge-density wave.  This can be seen in Fig.~\ref{fig:densityallVs}
where we show the time evolution of the particle density in the chain
for $V_s=V_c=V_{cs}=1,2,3,4$ and $\gamma=J=1$. As for $V_c=V_{cs}=0$,
$V_s\neq 0$, due to the large initial energy, $E_{\rm ini}=(L-1)V_s$,
of the system the number of accessible states is severely restricted.
Therefore larger interactions, $V\geq 3$, not only lead to a faster
relaxation but also suppress the particle density in the chain. 

\begin{figure}[htbp]
\begin{center}
\includegraphics*[width=0.7\textwidth]{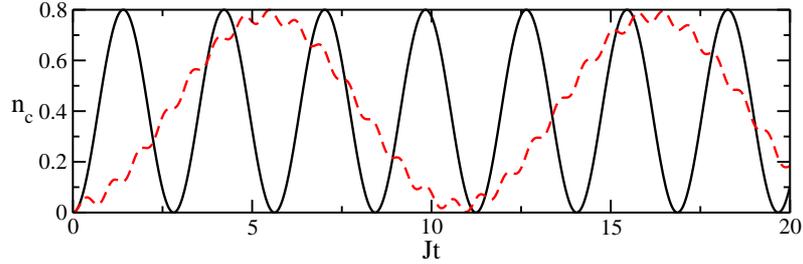}
\end{center}
\caption{Particle density in the chain $n_c(t)$ 
for $L=2$ lattice sites and $\gamma/J=1$
for the Hermitian time evolution.
We show exact results for non-interacting and interacting fermions: 
$V_c=V_s=V_{cs}=0$ (solid line),
$V_c=V_{cs}=V_s=8$ (dashed line).\label{FigL2}}
\end{figure}

In the limit where $V_s=V_c=V_{cs}$ is much larger than the bandwidth,
only the two states with the largest potential energies are reachable
due to energy conservation: a full bath or a full chain. In general,
the system can tunnel between these two possible states.  For a large
system size, however, the probability of all particles coherently
tunneling is very small and therefore the associated timescale is very
long. If we reduce the system size to $L=2$, on the other hand, the
oscillations between these states occur on a relatively short
timescale as can be seen in Fig.~\ref{FigL2}. For $L=2$ the
Hamiltonian can be exactly diagonalized for arbitrary interactions and
we show here exact results.

\begin{figure}[htbp]
\begin{center}
\includegraphics*[width=0.72\textwidth]{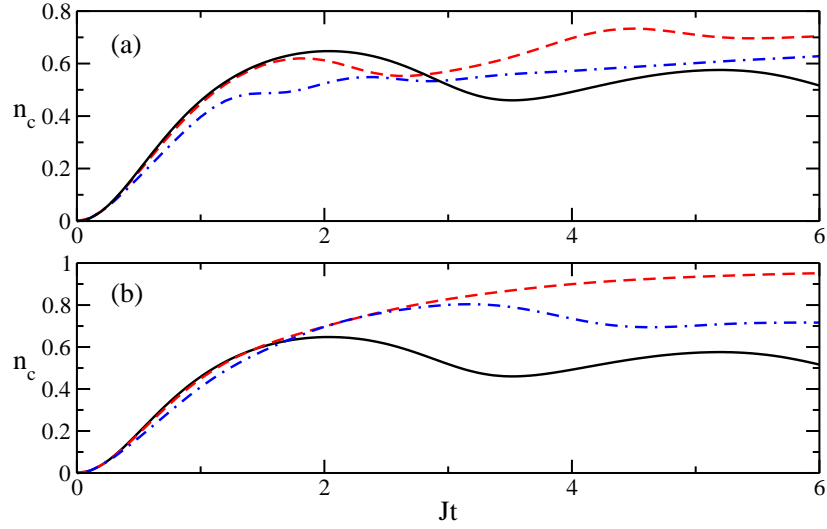}
\end{center}
\caption{Particle density in the chain $n_c(t)$ 
for $L=50$ lattice sites and $\gamma_{\rm in}/J=1$
for the ratchet time evolution.
We show $t$-DMRG results for non-interacting fermions 
($V_c=V_s=V_{cs}=0$, solid lines) in comparison
with interacting fermions;
(a) $V_c=V_{cs}=2, V_s=0$ (dashed line) and $V_c=V_{cs}=4, V_s=0$
(dot-dashed line);
(b) $V_c=V_{cs}=0, V_s=2$ (dashed line) 
and $V_c=V_{cs}=0, V_s=4$ (dot-dashed line).\label{Fig6}}
\end{figure}

Next, we consider the ratchet case with interactions.  Note that time
evolution is limited to $Jt\leq 6$ so that we cannot follow the time
evolution for too long.  In Fig.~\ref{Fig6} we show $t$-DMRG results
for the ratchet case with $\gamma_{\rm in}/J=1$.  Again, the
interactions damp the fluctuations and, for the specific cases
considered here.  lead to an increase of the average particle number
in the chain.

\subsection{Single-particle density matrix}
\label{sub:akk-interaction}

We will concentrate on the Hermitian time evolution for which $t$-DMRG
data are available for short and intermediate times.  In the
interacting case, scattering processes between fermions provide an
efficient mechanism for the relaxation of the single-particle density
matrix. Therefore, we expect that $A_{k,k}(Jt\gg 1)$ becomes
independent of time.  Note, however, that this distribution may, but
need not, correspond to a thermal distribution for an appropriately
chosen ensemble at a temperature determined by $\langle
\hat{H}\rangle_{\textrm{thermal}}\equiv\langle
\hat{H}\rangle_{\textrm{initial}}$.  The interesting question whether
or not this thermal state is reached will be addressed in a
forthcoming publication~\cite{InPrep}.

\begin{figure}[htbp]
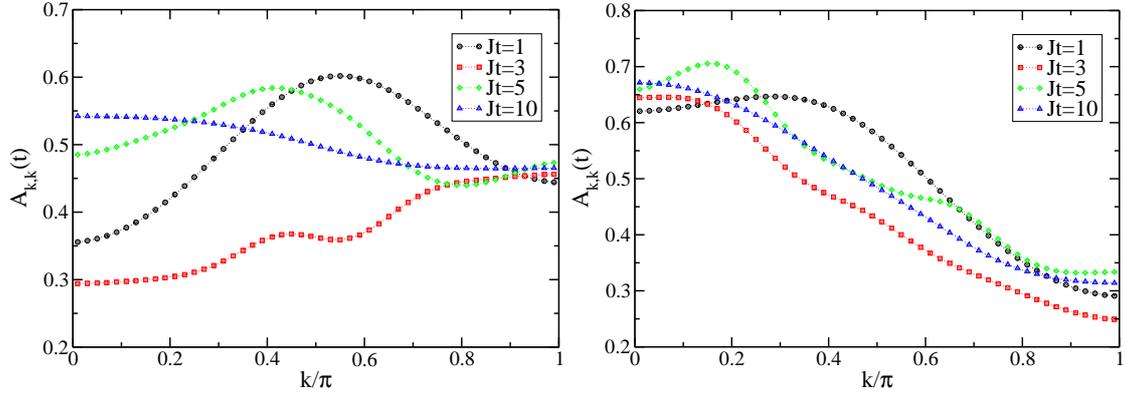

\begin{center}
\includegraphics*[width=0.49\textwidth]{L50_Akk_hermitian_interacting_V_V1_2.eps}
\includegraphics*[width=0.49\textwidth]{L50_Akk_hermitian_interacting_V2_2.eps}
\end{center}
\caption{Single-particle density matrix $A_{k,k}(t)$ for the Hermitian time
evolution with $\gamma/J=1$ for a chain of length $L=50$ with interaction
parameters $V_c=V_{cs}=2$, $V_s=0$ (left)
and $V_c=V_{cs}=0$, $V_s=2$ (right).\label{Fig10}}
\end{figure}

In Fig.~\ref{Fig10} we show results for interaction parameters
$V_c=V_{cs}=2$, $V_s=0$ and $V_c=V_{cs}=0$, $V_s=2$. For
$V_c=V_{cs}=2$, $V_s=0$ at short times, the single-particle density
matrix is still very close to the non-interacting case shown in
Fig.~\ref{Fig7}. At longer times, the interactions lead to a rapid
damping of the oscillations and, already at $Jt=10$, the distribution
in both cases becomes almost time-independent. For $V_c=V_{cs}=2$,
$V_s=0$ the distribution is very flat $A_{k,k}(t\gg 1)\simeq 1/2$.
This observation supports our interpretation of the data shown in
Fig.~\ref{Fig5}a: for strong interactions, the fermions tend to form a
Wigner lattice of alternatingly occupied chain and source sites which
corresponds to a completely flat single-particle density matrix. For
$V_c=V_{cs}=0$, $V_s=2$, on the other hand, the distribution at long
times resembles more a thermal Fermi distribution. Here the loss of
potential energy due to particles leaving the bath has to be
compensated by a gain in kinetic energy.  For the interaction strength
considered, approximately half of the particles enter the chain in the
long-time limit, see Fig.~\ref{Fig5}b. These particles are almost free
thus explaining why the long-time distribution shown in
Fig.~\ref{Fig10} is qualitatively similar to a thermalized free
fermion distribution.

In contrast to the non-interacting case shown in Fig.~\ref{Fig7},
$A_{k,k}(t)$ is a smooth function of the pseudo-momentum~$k$.  Despite
the fact that our $t$-DMRG cannot go much beyond $Jt=10$, our data
suggest that the electron-electron interactions lead to a fast
relaxation of the single-particle density matrix to its value in the
long-time limit.

\subsection{Current density}
\label{subsec:current-interaction}

\begin{figure}[hb]
\begin{center}
\includegraphics*[width=0.8\textwidth]{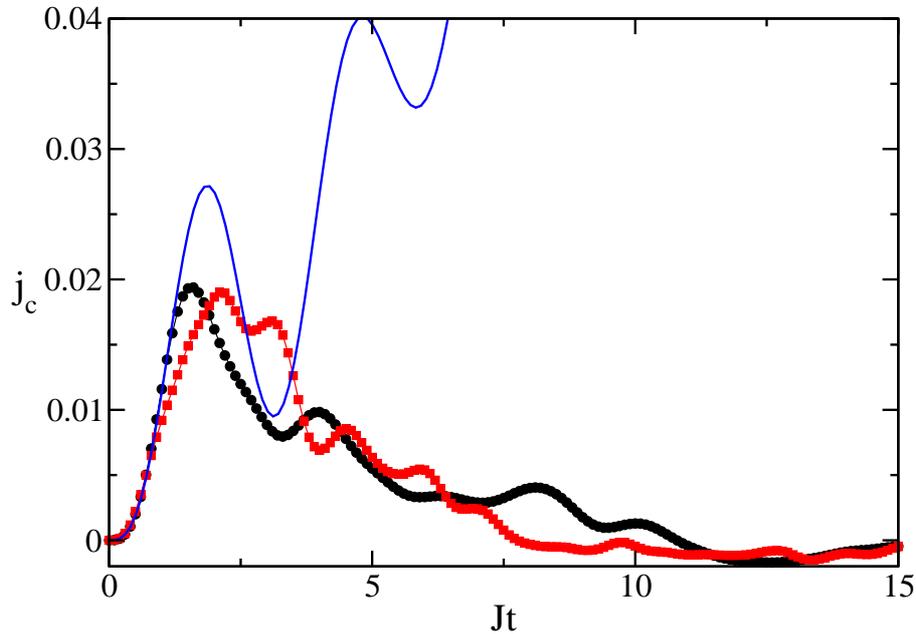}
\end{center}
\caption{Current density $j_c(t)$ 
for interacting electrons for $\gamma_l=J\exp(-l/L)$
in the Hermitian time evolution ($J=1$) for $L=50$ sites
with $V_c=V_{cs}=2, V_s=0$ (circles) and $V_c=V_{cs}=4, V_s=0$
(squares). For comparison we also show the result for noninteracting
fermions (full line).
\label{fig:interacting-currents}}
\end{figure}

For non-interacting particles,
the current decays to zero algebraically in time.
The reason is the dephasing of the $k$-modes.
In the presence of interactions, the current not only dephases but 
decays (exponentially) quickly. 
In Fig.~\ref{fig:interacting-currents} we show the result for the 
Hermitian time evolution for $L=50$ sites and two sets of interaction 
parameters. The fast decay of the current is clearly observed.

As discussed in Sect.~\ref{subsub:inhom-Hermitian_coupling},
the inhomogeneous injection of non-interacting fermions
results in a swapping of particles between the chain boundaries. 
As seen in Fig.~\ref{fig:non-interacting-currents}, the 
current density increases to a maximal value of 
$j_{c,{\rm max}}^{\rm non-int}\approx 0.1$
at $Jt\approx 50$, oscillates with period $T\approx L$,
and decays to zero algebraically slowly.
For interacting fermions, however, the current density reaches its
maximum of $j_{c,{\rm max}}^{\rm int}\approx 0.02$
at $Jt\approx 5$, and rapidly decays to very small values already at 
$Jt\approx 10$. Apparently, the interactions prevent the buildup
of a coherent mesoscopic particle wave package, and only local transient
current fluctuations can be seen. Note that this effect
is most prominent in one dimension where particles can scatter 
forward and backward only. In addition, we have chosen quite sizable
interactions to make the rapid relaxation visible 
within our restricted time interval in $t$-DMRG.
For a more detailed analysis, longer times must be studied.
This is beyond our numerical capabilities.

\section{Summary and Conclusions}
\label{sec:penultimate}

In this work, we studied the injection of spinless fermions 
from filled reservoir sites into an empty chain on which
the particles can move between neighboring sites.
We treated the cases of homogeneous and inhomogeneous
hybridizations between the source and chain sites
under a Hermitian time evolution and for the quantum ratchet case 
where the chain particles cannot return to the source sites.
For non-interacting electrons, we provide analytical expressions 
as a function of time for 
the particle density, the single-particle density matrix, 
and the current density. In the presence of interactions,
we use the $t$-DMRG algorithm to describe the time evolution
of observables for short to medium time scales.

The Hermitian time evolution and the ratchet time evolution both show
some qualitatively similar effects which is quite surprising at first.
Despite the fact that the particles cannot return from the chain to
the source sites in the ratchet case, their injection remains a
coherent quantum-mechanical process.  Therefore, the well-known
quantum-mechanical phenomena such as constructive and destructive
interference and Pauli blocking are seen in our observables.  In
particular, there are in both cases oscillations around the long-time
average of the chain filling.  Naturally, the results differ
quantitatively. For example, the chain filling (transient current) is
larger (smaller) for the ratchet case than for the Hermitian case for
the same parameter set.

The time evolution of non-interacting particles is characterized
by the absence of relaxation so that fairly large dephasing times, $Jt>10^3$, 
dominate the time evolution of our quantities. For large systems,
the dephasing ultimately leads to steady-state expressions for
the particle and current densities but the single-particle density matrix
reveals that the spectrum is that of a single-particle Hamiltonian.
For a homogeneous coupling, the system consists of independent two-level
systems and the single-particle density matrix reflects the
beating of these two-level systems. For inhomogeneous hybridizations,
the beating of the diagonal terms $A_{k,k}(t)$ is 
less pronounced because $k$~is no longer a good quantum number.
The inhomogeneous coupling between chain and source sites
generates non-diagonal elements in the single-particle density matrix 
which, however, are small and dephase in time. They give rise to 
the coherent motion of a spreading particle wave packet between the chain ends
which results in a transient oscillating current density.

When we introduce substantial interactions between the particles, the
picture changes qualitatively.  Depending on the type of
density-density interactions considered, the system starts with
different initial energies. Since the energy is conserved during time
evolution, only parts of the Hilbert space are accessible.
Interactions can therefore increase or decrease the particle density
in the chain compared to the non-interacting case. Furthermore, the
single-particle density matrix in the long-time limit can show a
completely flat distribution in cases where the potential energy
dominates, or become similar to a thermalized free fermion
distribution when the kinetic energy plays the dominant role. Most
importantly, instead of a slow dephasing or beating phenomena, we
observe a rapid relaxation to stationary values on time scales
$Jt=10$, not only for the particle and current densities but also for
the single-particle density matrix.  This shows the importance of
interactions for a fast relaxation for all physical quantities of
interest. In how far the obtained time-independent results do coincide
with those predicted by thermodynamic considerations is a question we
are planning to address in more detail elsewhere~\cite{InPrep}.

% \section{Conclusions}
% \label{sec:final}

%\subsection{Change marks}

%Please use for changes requested 
%by the referee the following colour change option:
%\begin{changed}
%  This is a text snippet marked as \emph{changed}. 
%\end{changed}

\begin{acknowledgement}
J.S.\ acknowledges support by the MAINZ (MATCOR) school of excellence
and the DFG via the SFB/Transregio 49.
\end{acknowledgement}

\appendix
\section{Single-particle density matrix for non-interacting fermions
(inhomogeneous ratchet case)}
\label{app:D}
\subsection{State}
As seen from eq.~(\ref{eq:psi-t-inhomratchet}), 
the time-dependent state is given by
\begin{align}
|\psi(t)\rangle
= \prod_{k=1}^L\left( 
1+ g_k(t)\hat{B}_k\hat{a}_k^{\dagger}\right)
|\psi(0)\rangle \; , 
\end{align}
where $|\psi(0)\rangle=\prod\limits_{l=1}^L \hat{s}_l^{\dagger}|{\rm vac}\rangle$ 
and $g_k(t)=(1-e^{-\rmi \varepsilon_k t})/\varepsilon_k$.
We rewrite the product as
\begin{subequations}
\begin{eqnarray}
 |\psi(t)\rangle&= &\Bigl[ 1  \label{eq:firstterm}\\
 && +\sum_{k} g_k \hat{B}_k\hat{a}_k^{\dagger}\label{eq:secondterm}\\
 && +\sum_{k_1<k_2}g_{k_1}g_{k_2} \hat{B}_{k_1}\hat{a}_{k_1}^{\dagger}
\hat{B}_{k_2}\hat{a}_{k_2}^{\dagger}\\
 && +\sum_{k_1<k_2<k_3}g_{k_1}g_{k_2}g_{k3} 
\hat{B}_{k_1}\hat{a}_{k_1}^{\dagger}
\hat{B}_{k_2}\hat{a}_{k_2}^{\dagger}
\hat{B}_{k_3}\hat{a}_{k_3}^{\dagger}\\
 && +\dots \Bigr] |\psi(0)\rangle \; . \nonumber
\end{eqnarray}
\end{subequations}
The first term, eq.~(\ref{eq:firstterm}), 
describes a state where there is no particle on the chain, 
the second term, eq.~(\ref{eq:secondterm}) 
describes the situation with one particle on the chain, 
and so on.
\subsection{Numerator}
In the density matrix $A_{p,q}$, the operator $\hat{a}_p^{\dagger}\hat{a}_q$ 
does not change the number of particles on the chain. Therefore,
\begin{subequations}\begin{eqnarray}
\label{eq:appdm1}
\langle \psi(t)| \hat{a}_p^{\dagger}\hat{a}_q | \psi(t)\rangle &=& 
g_q g_p^* \langle \hat{B}_p^{\dagger} \hat{B}_q\rangle_0\\
&&+ \sum_{k\neq (p,q)} \frac{g_q g_k}{2}
\sum_{k_1\neq k_2} g_{k_1}^*g_{k_2}^*
\langle 
\hat{a}_{k_2}\hat{B}_{k_2}^{\dagger}
\hat{a}_{k_1}\hat{B}_{k_1}^{\dagger}
\hat{B}_{k}\hat{a}_{k}^{\dagger}
\hat{B}_{q}\hat{a}_{p}^{\dagger}
\rangle_0 
\label{eq:2ord}\\
& &+ \sum_{k_1<k_2 \neq (q,p)}\frac{g_q g_{k_1} g_{k_2}}{6}
\begin{array}[t]{@{}l@{}}
\sum\limits_{\substack{s_1,s_2,s_3\\s_1\neq s_2\neq s_3\neq s_1}} 
g_{s_1}^* g_{s_2}^* g_{s_3}^* \\[18pt]
\phantom{\sum_A^B}
\times \langle 
\hat{a}_{s_3}\hat{B}_{s_3}^{\dagger}
\hat{a}_{s_2}\hat{B}_{s_2}^{\dagger}
\hat{a}_{s_1}\hat{B}_{s_1}^{\dagger}
\hat{B}_{k_1}\hat{a}_{k_1}^{\dagger}
\hat{B}_{k_2}\hat{a}_{k_2}^{\dagger}
\hat{B}_{q}\hat{a}_{p}^{\dagger}
\rangle_0
\end{array}
\label{eq:3ord}\\
& &+\dots\; .\nonumber
\end{eqnarray}\end{subequations}
We can factor out the expectation value for the chain operators 
and use Wick's theorem~\cite{Fetter} so that we can identify some
of the momenta.
Relabeling some indices leads to the following expressions for 
the second and third term of the numerator,
\begin{eqnarray}
 \text{(\ref{eq:2ord})}&=&\sum_{k\neq(p,q)} g_q g_k g_k^* g_p^* 
\langle
\hat{B}_{p}^{\dagger}\hat{B}_{k}^{\dagger}\hat{B}_{k}\hat{B}_{q}
\rangle_0 \; , \label{eq:2ord-new}\\
\text{(\ref{eq:3ord})}&=&\sum_{\substack{k_1<k_2\\ k_1,k_2\neq (q,p)}} 
g_q g_p^* |g_{k_1}|^2 |g_{k_2}|^2 
\langle
\hat{B}_{p}^{\dagger}\hat{B}_{k_2}^{\dagger}\hat{B}_{k_1}^{\dagger}
\hat{B}_{k_1}\hat{B}_{k_2}\hat{B}_{q}
\rangle_0 \; .
\label{eq:3ord-new}
\end{eqnarray}
The expectation values are calculated using Wick's theorem. 
To this end we need
$\langle \hat{B}_{p}^{\dagger}\hat{B}_{q} \rangle_0
= H(p,q)$, see eq.~(\ref{eq:defHPQ}).
The expectation values in eqs.~(\ref{eq:2ord-new}) and~(\ref{eq:3ord-new})
can be expressed as determinants,
\begin{eqnarray}
\langle
\hat{B}_{p}^{\dagger}\hat{B}_{k}^{\dagger}\hat{B}_{k}\hat{B}_{q}
\rangle_0 
&=& 
\left|
\begin{array}{cc} H(p,q)&H(p,k)\\H(k,q)&H(k,k)
\end{array}\right|
\\[6pt]
\langle
\hat{B}_{p}^{\dagger}\hat{B}_{k_2}^{\dagger}\hat{B}_{k_1}^{\dagger}
\hat{B}_{k_1}\hat{B}_{k_2}\hat{B}_{q}
\rangle_0 &=&
\left|
\begin{array}{ccc}H(p,q)&H(p,k_1)&H(p,k_2) \\ 
H(k_1,q)&H(k_1,k_1) & H(k_1,k_2) \\ 
H(k_2,q) &H(k_2,k_1)&H(k_2,k_2) 
\end{array} \right|\; .
\end{eqnarray}
This leads to the result 
\begin{eqnarray}
\label{eq:appden2}
 \langle \psi(t)|
\hat{a}_p^{\dagger}\hat{a}_q | \psi(t)\rangle &=&
g_q g_p^* \Biggl[ 
H(p,q) + \sum_k |g_k|^2  
\left|\begin{array}{cc} 
H(p,q)&H(p,k)\\
H(k,q)&H(k,k)
\end{array}\right| 
\\[6pt]
&&+ \frac{1}{2} \sum_{k_1,k_2} |g_{k_1}|^2 |g_{k_2}|^2 
\left|
\begin{array}{ccc}
H(p,q)&H(p,k_1)&H(p,k_2) \\ 
H(k_1,q)&H(k_1,k_1) & H(k_1,k_2) \\ 
H(k_2,q) &H(k_2,k_1)&H(k_2,k_2) 
\end{array} \right|
+ \ldots \Biggr]\; .\nonumber
\end{eqnarray}
We were allowed to drop the restriction on the summations 
due to the properties of the determinants.

\subsection{Self-energy}
As in all perturbative calculations, we can translate the 
terms in~(\ref{eq:appden2}) into diagrams~\cite{Fetter}
with `internal vertices'~$k$ of strength $f_k=|g_k|^2$
and lines between vertices $k_1$ and $k_2$ representing the factors 
$H(k_1,k_2)$. In the thermodynamic limit, $L \to\infty$, 
the linked-cluster theorem applies~\cite{Fetter}, i.e., the
denominator $\langle \psi(t) | \psi(t)\rangle$ in $A_{p,q}$
cancels the unconnected diagrams in the numerator.
In $A_{p,q}$ we have to sum over all diagrams which connect 
the external vertex~$p$ with the external vertex~$q$. We find
\begin{equation}
\label{eq:densitymatrixdia}
 A_{p,q} = g_q g_p^* \left[  H(p,q) - \sum_k H(p,k) f_k H(k,q)
+\sum_{k_1,k_2} H(p,k_1)f_{k_1}H(k_1,k_2)f_{k_2}H(k_2,q) \mp \ldots \right]
\; .
\end{equation}
We introduce the self-energy $\Sigma(a,b)$~\cite{Fetter} and prove
eq.~(\ref{eq:solveApq}),
\begin{equation}
 A_{p,q}=g_q g_p^* \left( H(p,q) + 
\sum_{k_1,k_2} H(p,k_1) \Sigma(k_1,k_2) H(k_2,q) \right)\; ,
\end{equation}
where the self-energy entry $\Sigma(a,b)$ is obtained by summing 
all possible paths from the internal vertex $a$ to the internal vertex~$b$ 
with altogether~$n$ internal vertices ($n=1,2,\ldots$)
and $(n-1)$ lines between them,
\begin{eqnarray}
\Sigma(a,b)&=&-\delta_{a,b} f_a + f_a \Bigl[ H(a,b)f_b 
- \sum_{k}H(a,k) f_k H(k,b)f_b   \nonumber \\
&& + \sum_{k_1,k_2}  \nonumber 
H(a,k_1) f_{k_1} H(k_1,k_2) f_{k_2} H(k_2,b)f_b  \mp \ldots \Bigr]\\
&=&-\delta_{a,b} f_a - f_a \sum_{k} H(a,k) \Sigma(k,b)\; . \label{eq:sigma}
\end{eqnarray}
The solution of this matrix equation provides $\mathbf{\Sigma}$, 
see eq.~(\ref{eq:Sigmasolution}).

%%\def\bstname{adp}
%%\bibliographystyle{adp}
%%\bibliography{injection}
\providecommand{\WileyBibTextsc}{}
\let\textsc\WileyBibTextsc
\providecommand{\othercit}{}
\providecommand{\jr}[1]{#1}
\providecommand{\etal}{~et~al.}

\end{document}